\documentclass[article,10pt,a5paper,oneside,cyremdash]{ncc}
\FromMargins[h]{2.4cm}{1.6cm}{.8cm}{1.6cm}
\usepackage{cmap}
\usepackage[boldsans]{concmath}
\usepackage{amssymb,fancyvrb,cite,icomma,indentfirst}
\fvset{fontsize=\footnotesize,frame=single,framesep=.4em,
       xleftmargin=0.2em,xrightmargin=0.2em}
\usepackage[utf8]{inputenc}
\usepackage[T1,T2A]{fontenc}
\usepackage[english,russian]{babel}

\newcommand{\hm}[1]{#1\nobreak\discretionary{}{\hbox{\ensuremath{#1}}}{}}

\usepackage[pdftex,unicode]{hyperref}
\hypersetup{unicode=true,
    pdfauthor={P.V. Moskalev, K.V. Grebennikov, V.V. Shitov},
    pdfsubject={Statistical estimation of percolation cluster parameters},
    pdftitle={P.V. Moskalev et al. Statistical estimation of percolation cluster parameters},
    pdfkeywords={mathematical modeling, Monte Carlo methods, site percolation, percolation cluster, mass fractal dimension} }

\begin{document}

\begin{titlepage}
\English
\begin{flushleft}\bfseries\Large
Statistical estimation of \\percolation cluster parameters
\end{flushleft}
\begin{flushleft}\bfseries
P.V. Moskalev$^\dag$, K.V. Grebennikov$^\ddag$, V.V. Shitov$^\ddag$
\end{flushleft}
\small
\begin{itemize}
\item[$^\dag$]
Voronezh State Agricultural University after K.D. Glinki \\(moskalefff@gmail.com)
\item[$^\ddag$]
Voronezh State Technological Academy (svw@list.ru)
\end{itemize}
\begin{flushleft}
Received: April 18, 2011; in final form: May ??, 2011
\end{flushleft}

\paragraph{Abstract.} In this paper we study statistical methods of parameters estimation of the site percolation model. Advantages of the proposed method is demonstrated for the computing of the confidence interval of mass fractal dimension of a percolation clusters sampling, formed by the Monte Carlo method.


\clearpage
\Russian

\begin{flushleft}\bfseries
УДК 519.676
\end{flushleft}
\begin{flushleft}\bfseries\Large
Статистическое оценивание \\характеристик перколяционного кластера
\end{flushleft}
\begin{flushleft}\bfseries
П.В. Москалев$^\dag$, К.В. Гребенников$^\ddag$, В.В. Шитов$^\ddag$
\end{flushleft}
\small
\begin{itemize}
\item[$^\dag$]
Воронежский государственный аграрный университет \\имени К.Д. Глинки (moskalefff@gmail.com)
\item[$^\ddag$]
Воронежская государственная технологическая академия \\(svw@list.ru)
\end{itemize}
\begin{flushleft}
Поступила в Редакцию: 18 апреля 2011 г. \\
В окончательной редакции: ?? мая 2011 г.
\end{flushleft}

\paragraph{Аннотация.}
В данной работе рассматриваются статистические методы оценивания характеристик математической модели перколяции узлов. Преимущества предлагаемой методики демонстрируются на примере вычисления интервальной оценки массовой фрактальной размерности по выборочной совокупности перколяционных кластеров, формируемой с помощью метода статистических испытаний.

\paragraph{Ключевые слова:} 
математическое моделирование, метод статистических испытаний, перколяция узлов, перколяционный кластер, массовая фрактальная размерность.

\tableofcontents

\normalsize

\end{titlepage}

\section{Введение}

В классическом определении критического явления основной акцент делается на описании поведения вещества в окрестности точки фазового перехода. При этом, если объем новой фазы увеличивается постепенно, то говорят о фазовом переходе первого рода, если же сосуществование двух фаз исключено и процесс имеет скачкообразный характер, то речь идёт о фазовом переходе второго рода \cite{moskaleff.2009.06}. С более общих позиций критические явления характеризуют процесс перехода некоторой системы из одного устойчивого состояния в другое, а о принадлежности этого перехода к первому или второму роду можно говорить в зависимости от того, как быстро он происходит. Одной из важных особенностей фазовых переходов второго рода является связь процесса формирования новой фазы с индивидуальными флуктуациями, которые в окрестности критической точки приобретают коррелированный характер с радиусом корреляции, сопоставимым с размером системы.

Математическая теория перколяции занимается изучением так называемых геометрических фазовых переходов на взвешенных ориентированных однородных графах (в простейшем случае на решётках). Основной особенностью перколяционных решёток является зависимость весовых коэффициентов вершин и/или дуг от пространственного и вероятностного распределения псевдослучайных чисел. В соответствии с принятой в теории перколяции терминологией, ребра графа называют связями, его вершины\--- узлами, а основными объектами изучения являются связанные подмножества узлов решётки, называемые кластерами. В подобной системе фазовый переход второго рода соответствует появлению перколяционного кластера, размер которого потенциально бесконечен. В ограниченной области такой кластер идентифицируют по размеру, соответствующему размеру всей перколяционной решётки. 

Для реализации всех описанных в данной статье алгоритмов была использована свободная система статистической обработки данных и программирования GNU/R \cite{r-project.org}, возникшая в 1993 году как свободная альтернатива языка S, разработанного в конце 1970-х годов в компании Bell Labs специально для решения задач вычислительной статистики. По замыслу ее создателей, система R должна была обладать легко расширяемой модульной архитектурой при сохранении быстродействия, присущего программам, написанным на FORTRAN. 

В настоящее время сборки системы R существуют для наиболее распространённых семейств операционных систем: Apple Mac OS X, GNU/Linux и Microsoft Windows, а в распределённых хранилищах системы R по состоянию на середину апреля 2011 года доступны для свободной загрузки почти 3000 пакетов расширения, ориентированных на специфические задачи обработки данных, возникающие в финансовом анализе, генетике, экологии, фармацевтике и многих других прикладных областях \cite{cran.org}. Из учебной литературы на русском языке по использованию системы R можно указать на пособие \cite{statistics.with.r.2010}, вышедшее в прошлом году.

\section{Изотропная перколяция узлов}

Одна из простейших моделей некоррелированной перколяции получается на изотропной квадратной решётке $[0, 1]^2$ при взвешивании её узлов стандартной равномерно распределённой последовательностью псевдослучайных чисел $u_{xy} \sim \mathbf{U}[0, 1)$. Состав подмножества узлов кластера определяется итерационной процедурой сравнения псевдослучайных чисел $u_{xy} < p$, соответствующих узлам $(x, y)$ из единичной окрестности текущего подмножества $\{(x, y)\}$, с фиксированным числом $0 < p \leqslant 1$, соответствующим относительной доле достижимых узлов перколяционной решётки. Удовлетворяющие указанному неравенству узлы формируют на решётке кластеры, представляющие собой реализации псевдослучайного перколяционного процесса. 

\begin{figure}[hbt]
\centering
\includegraphics[width=.45\linewidth]{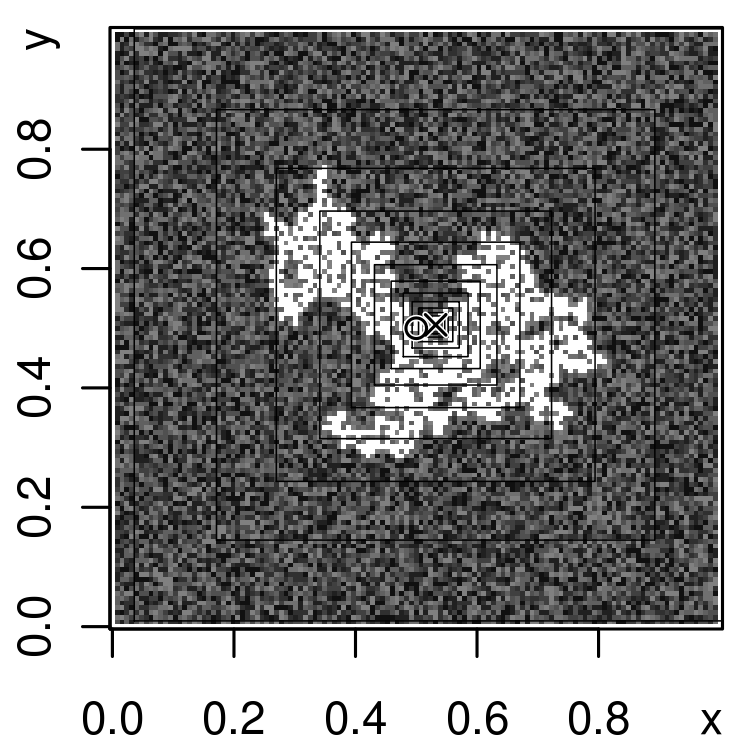}\quad
\includegraphics[width=.45\linewidth]{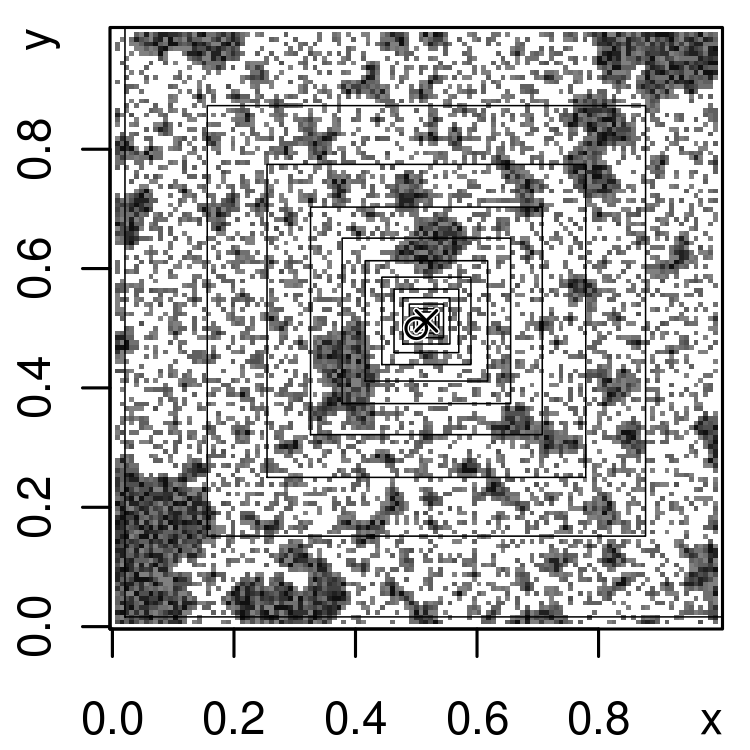}
\caption{Типичные реализации кластеров при перколяции из $(0,5; 0,5)$ на квадратной изотропной решётке при $p=p_c-0,02$ (слева) и $p=p_c+0,02$ (справа)}
\label{pic:clusters}
\end{figure}

На рис.\,\ref{pic:clusters} хорошо видно, что при доле достижимых узлов, меньшей некоторого критического значения $p\hm=p_c-0,02$, где $p_c\hm=0,592746\ldots$, перколяционный процесс является ограниченным, а при доле достижимых узлов, большей критического значения $p\hm=p_c+0,02$,\--- распространяется по всей решётке.
Реализация алгоритма построения подобных кластеров для плоской квадратной изотропной решётки показан в листинге 1.

\begin{table}[tbh]\small
\textbf{Листинг 1.} Построение кластера узлов на квадратной изотропной решётке
\begin{Verbatim}
psi20 <- function(x=129, p=0.592746, 
                  sst=(x*x+1)/2, trg=c(), label=2) {
    acc <- array(runif(x*x), c(x,x))
    acc[sst] <- label
    acc[c(1,x),] <- acc[,c(1,x)] <- label + 0.001
    repeat {
        acc[sst <- unique(c(
            sst[acc[sst - x] < p] - x,
            sst[acc[sst - 1] < p] - 1,
            sst[acc[sst + 1] < p] + 1,
            sst[acc[sst + x] < p] + x))] <- label
        if (length(sst) < 1) break }
return(acc) }
\end{Verbatim}
\end{table}

В качестве аргументов функции ``\texttt{psi20()}'' используются следующие переменные: ``\texttt{x}''\--- линейный размер решётки; ``\texttt{p}''\--- доля достижимых узлов решётки; ``\texttt{sst[]}'', ``\texttt{trg[]}''\--- векторы индексов для стартового и целевого подмножеств; ``\texttt{label}''\--- кластерная метка. В процессе выполнения функции ``\texttt{psi20()}'' формируется двумерный массив ``\texttt{acc[,]}'', изначально заполняемый псевдослучайными числами $u_{xy}$ со стандартным равномерным распределением, а элементы, соответствующие стартовому подмножеству, помечаются меткой ``\texttt{label}''. Затем, на базе цикла с постусловием организуется итерационный процесс в ходе которого узлы, принадлежащие единичной окрестности текущего стартового подмножества и удовлетворяющие условию $u_{xy} < p$, помечаются заданной меткой ``\texttt{label}'' и формируют новое стартовое подмножество для следующей итерации. Тогда, появление на очередной итерации вектора ``\texttt{sst[]}'' нулевой длины означает выход из цикла и возврат массива ``\texttt{acc[,]}'' в качестве результата функции ``\texttt{psi20()}''.

\section{Статистическое оценивание характеристик реализаций кластеров}

Статистически самоаффинным называют множество, допускающее разбиение на конечное число непересекающихся подмножеств, обладающих статистическими характеристиками, идентичными исходному множеству с точностью до некоторого аффинного преобразования. Одной из важнейших структурных характеристик подобного множества \cite{tarasevitch.2002} является массовая фрактальная размерность $d_{c1}$, определяемая методом наименьших квадратов для единичной реализации перколяционного кластера как коэффициент линейной регрессии логарифма числа узлов кластера $n_i$, покрываемых квадратом текущего размера $r_i$, к логарифму этого размера $(\ln r_i, \ln n_i)$:
\begin{equation}\label{eq:cls.reg}
\ln n_i = d_{c1} \ln r_i + d_{c0} + e_{ci}, \qquad
\sum\limits_{i=1}^{k} e_{ci}^2 \to \min,
\end{equation}
где $e_{ci}$\--- минимизируемые методом наименьших квадратов отклонения эмпирических точек от линии регрессии. Центром симметрии покрывающего множества является центр масс анализируемого кластера, определяемый по выборочным средним координатам его узлов. Для показанных на рис.\,\ref{pic:clusters} кластеров стартовое подмножество расположено в точке $(0,5; 0,5)$ и обозначено символом ``$\circ$'', а центры масс кластеров расположены в точках $(0,532; 0,506)$ и $(0,517; 0,512)$ и обозначены символами ``$\times$''.

\begin{figure}[hbt]
\centering
\includegraphics[width=.45\linewidth]{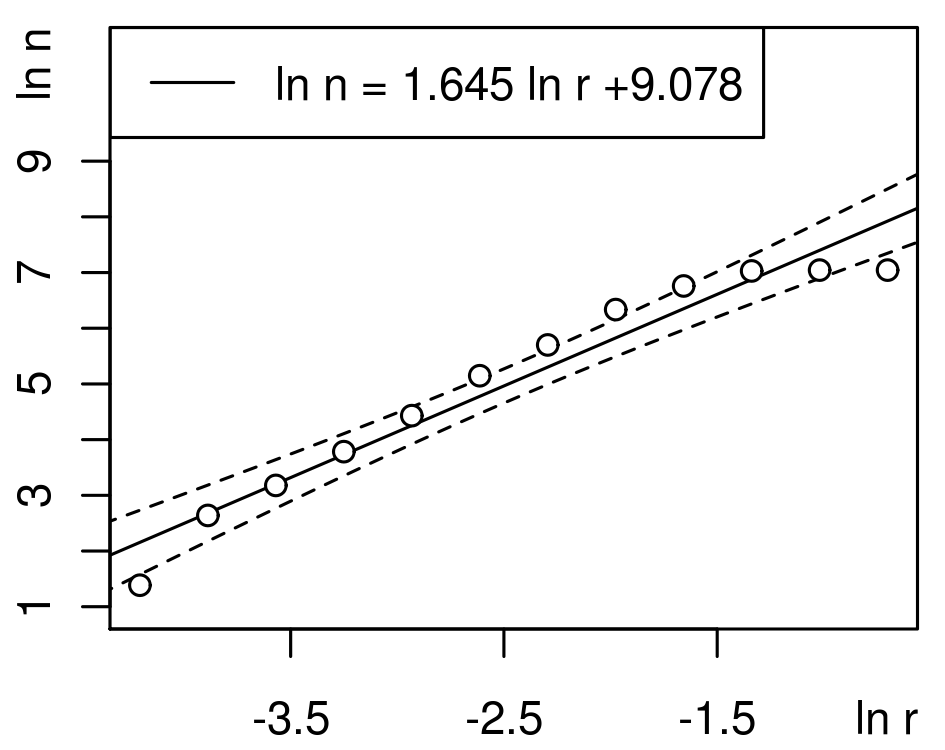}\quad
\includegraphics[width=.45\linewidth]{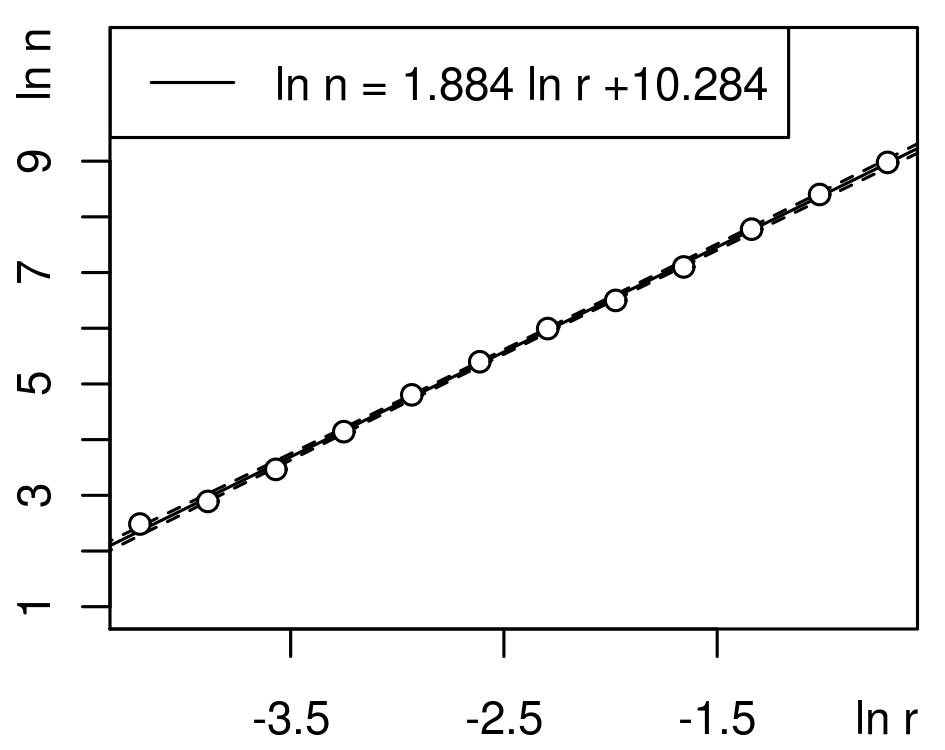}\\[1ex]
\includegraphics[width=.45\linewidth]{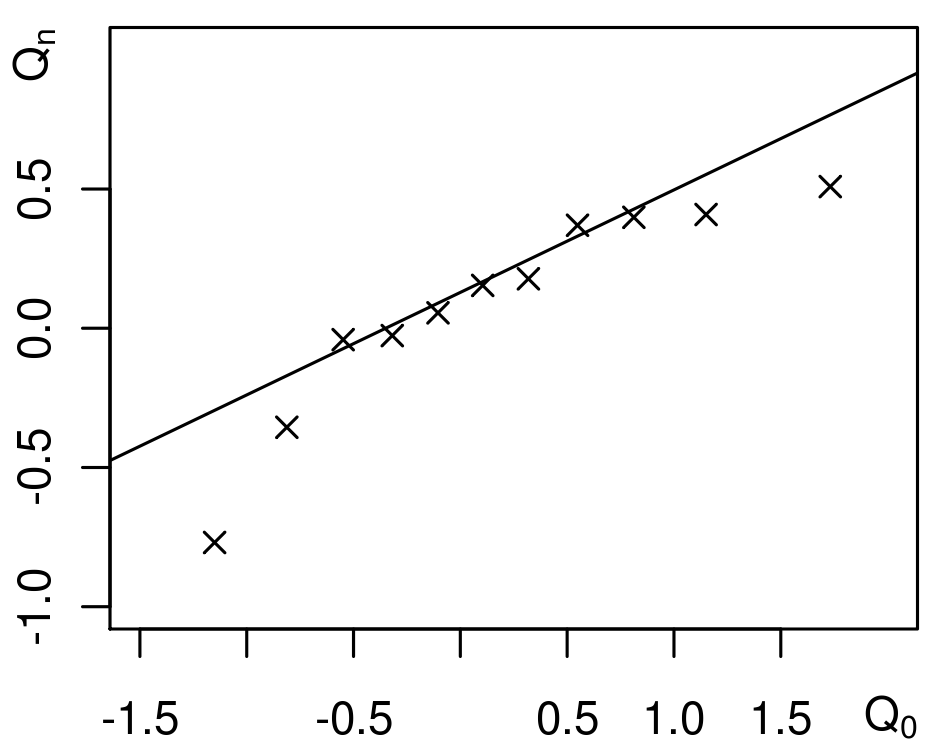}\quad
\includegraphics[width=.45\linewidth]{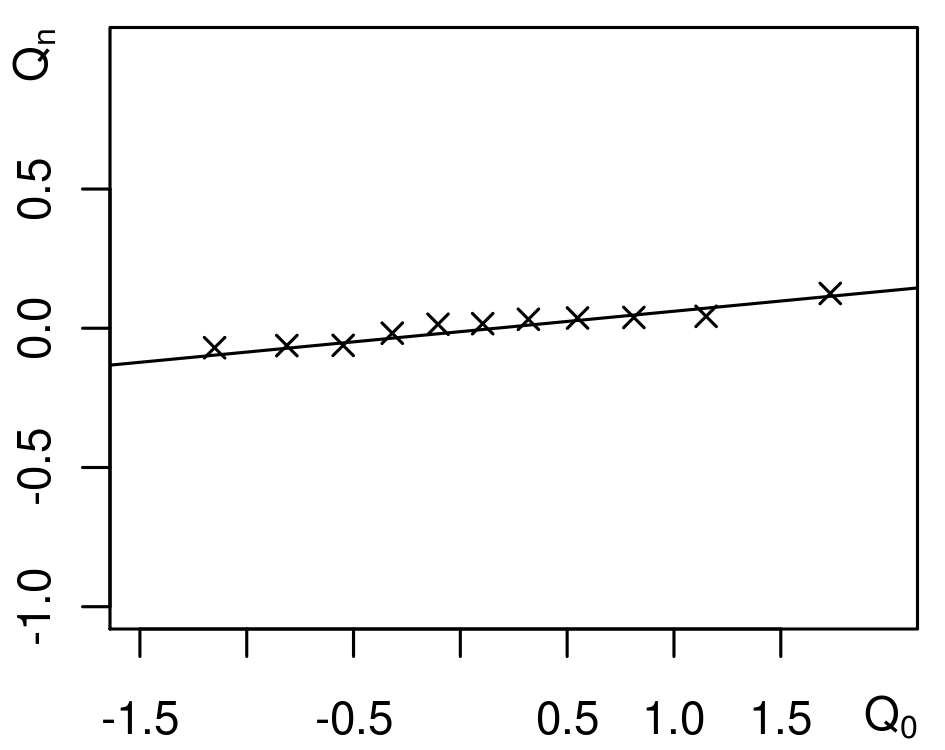}
\caption{Линейные модели (вверху) и Q\--Q графики остатков $e_{ci}$ (внизу) к оценке массовой фрактальной размерности $d_{c1}$ для показанных на рис.\,\ref{pic:clusters} реализаций кластеров при $p=p_c-0,02$ (слева) и $p=p_c+0,02$ (справа)}
\label{pic:cls.reg}
\end{figure}

Решение задачи линейной регрессии \eqref{eq:cls.reg} для статистического оценивания массовой фрактальной размерности $d_{c1}$ анализируемых кластеров показано на рис.\,\ref{pic:cls.reg}, а соответствующие $p_c-0,02$ и $p_c+0,02$ долям достижимых узлов точечные оценки $d_{c1}$ равны $1,645$ и $1,884$ при коэффициентах детерминации $R^2$, близких к единице: $0,940$ и $0,999$. Символами ``$\circ$'' на рис.\,\ref{pic:cls.reg} вверху показаны эмпирические точки $(\ln r_i, \ln n_i)$, сплошными линиями\--- линии регрессии, а штриховыми\--- их $0,95$\-/доверительные интервалы, методика построения которых описана, например, в \cite{statistics.with.r.2010}. 

Заметим, что из-за ограниченности перколяционного процесса при докритических значениях доли достижимых узлов $p=p_c-0,02$ средний радиус доверительного интервала для получаемой линии регрессии оказывается значительно более широким, а распределение отклонений эмпирических значений от теоретических, судя по всему, слабо соответствует гипотезе о нормальном распределении ошибок. Последний вывод подтверждается при сравнительном анализе квантиль\-/квантильных (Q\--Q) графиков остатков $e_{ci}$, показанных на рис.\,\ref{pic:cls.reg} внизу. Символами ``$\times$'' показаны соответствующие $e_{ci}$ теоретические и эмпирические квантили, а сплошной линией\--- теоретическая кривая нормального распределения остатков $F_0(e)$. На Q\--Q графиках хорошо видно, что при $p<p_c$ эмпирическое распределение остатков $e_{ci}$ будет сильно отличаться от нормального, что существенно снижает достоверность оценок фрактальной размерности реализаций кластеров $d_{c1}$ на этих режимах, не взирая на высокие значения коэффициента детерминации $R^2$.

\section{Статистическое оценивание характеристик выборки кластеров}

Из курса математической статистики хорошо известно, что повышение достоверности получаемых оценок $d_{c1}$ достигается за счёт построения репрезентативной выборочной совокупности, что приводит к повышению вычислительной сложности задачи, пропорциональному объёму используемой выборки. 

Как показывают наши недавние исследования \cite{moskaleff.2011.02}, характеристики перколяционного процесса зависят не только от размера и типа решётки, но и от взаимного расположения на решётке стартового и целевого подмножеств. Иными словами речь идёт о статистической связи распределения относительных частот узлов перколяционной решётки с характеристиками выборочной совокупности кластеров.

Тогда, для нахождения оценки массовой фрактальной размерности по выборке перколяционных кластеров можно воспользоваться методом, предложенным в несколько другом контексте в работе \cite{moskaleff.2009.06}. Этот метод использует относительные частоты узлов в качестве весовых коэффициентов, что позволяет находить статистическую оценку размерности перколяционного кластера сразу для всей их совокупности.

Применительно к данной задаче массовая фрактальная размерность выборки кластеров $d_{b1}$ будет определяться как коэффициент линейной регрессии логарифма относительных частот узлов $v_i$, покрываемых квадратом текущего размера $r_i$, к логарифму этого размера $(\ln r_i, \ln v_i)$:
\begin{equation}\label{eq:smp.reg}
\ln v_i = d_{b1} \ln r_i + d_{b0} + e_{bi}, \qquad
\sum\limits_{i=1}^{k} e_{bi}^2 \to \min,
\end{equation}
где $e_{bi}$\--- минимизируемые методом наименьших квадратов отклонения эмпирических точек от линии регрессии. Подобный переход вполне правомерен, так как при равных объёмах выборки средние значения абсолютных частот узлов перколяционной решётки $\langle ni\rangle$ соответствуют их относительным частотам $v_i$, что в пределе приводит к идентичности математических ожиданий оценок параметров $d_{c1}$ и $d_{b1}$ моделей \eqref{eq:cls.reg} и \eqref{eq:smp.reg}.

\begin{figure}[hbt]
\centering
\includegraphics[width=.45\linewidth]{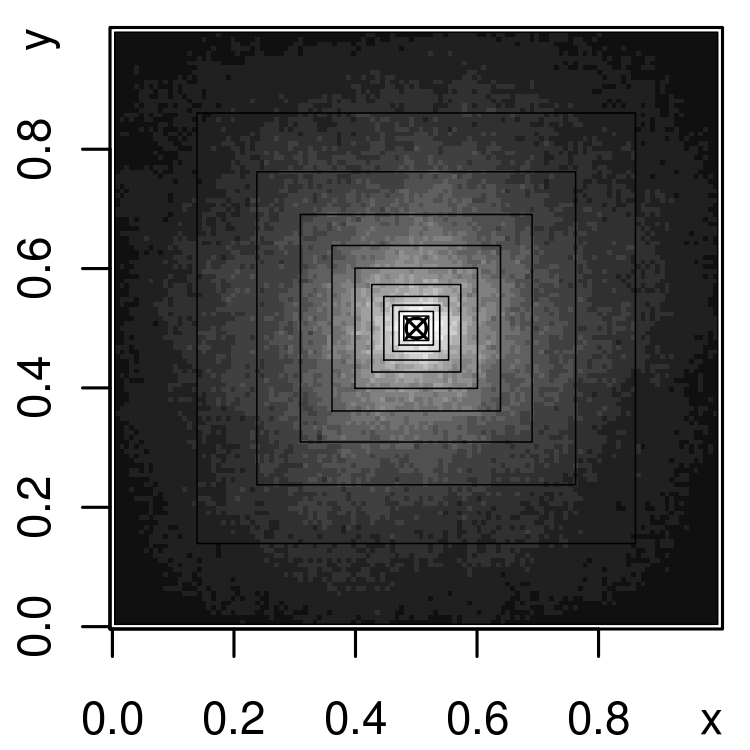}\quad
\includegraphics[width=.45\linewidth]{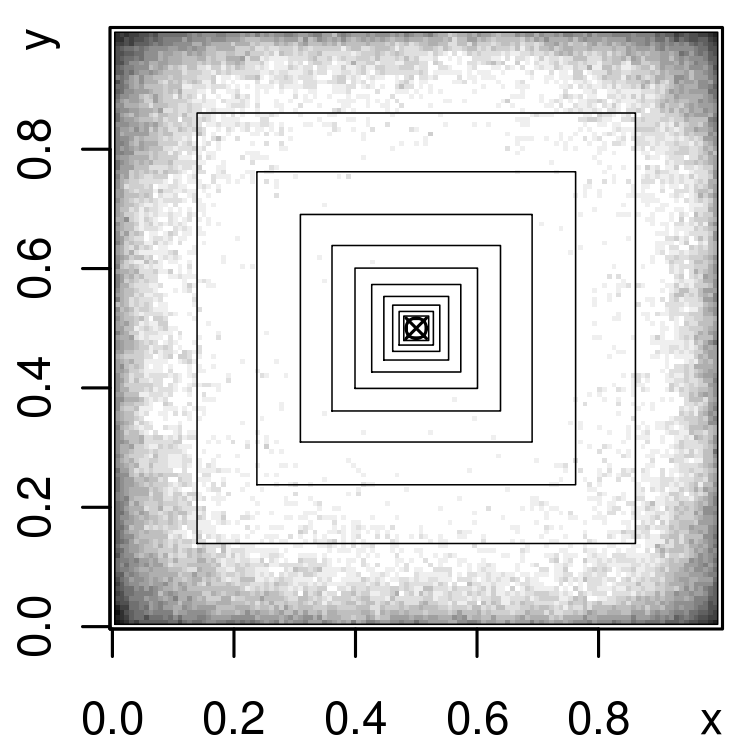}
\caption{Типичные выборки кластеров при перколяции из $(0,5; 0,5)$ на квадратной изотропной решётке при $p=p_c-0,02$ (слева) и $p=p_c+0,02$ (справа)}
\label{pic:samples}
\end{figure}

На рис.\,\ref{pic:samples} показаны распределения относительных частот узлов решётки $v_i$ для выборочных совокупностей объёмом $500$ кластеров, стартовое подмножество которых расположено в центральной точке области $(0,5; 0,5)$ и обозначено символом ``$\circ$''. С учётом изотропной постановки задачи центры симметрии покрывающих множеств, обозначенные на рис.\,\ref{pic:samples} символами ``$\times$'', будут совпадать со стартовым множеством.

\begin{figure}[hbt]
\centering
\includegraphics[width=.45\linewidth]{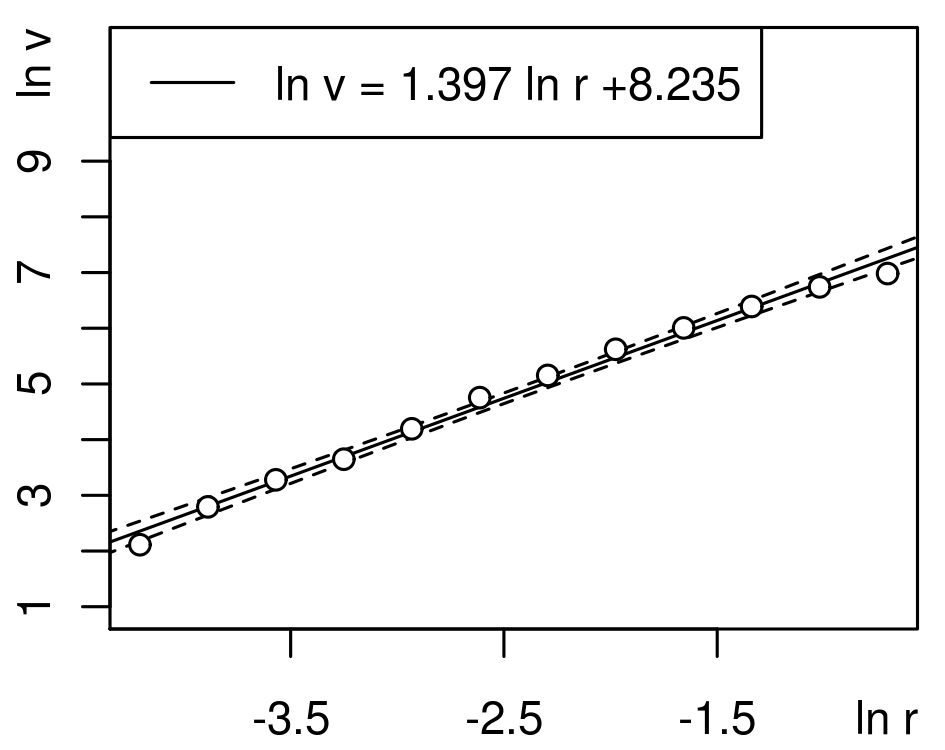}\quad
\includegraphics[width=.45\linewidth]{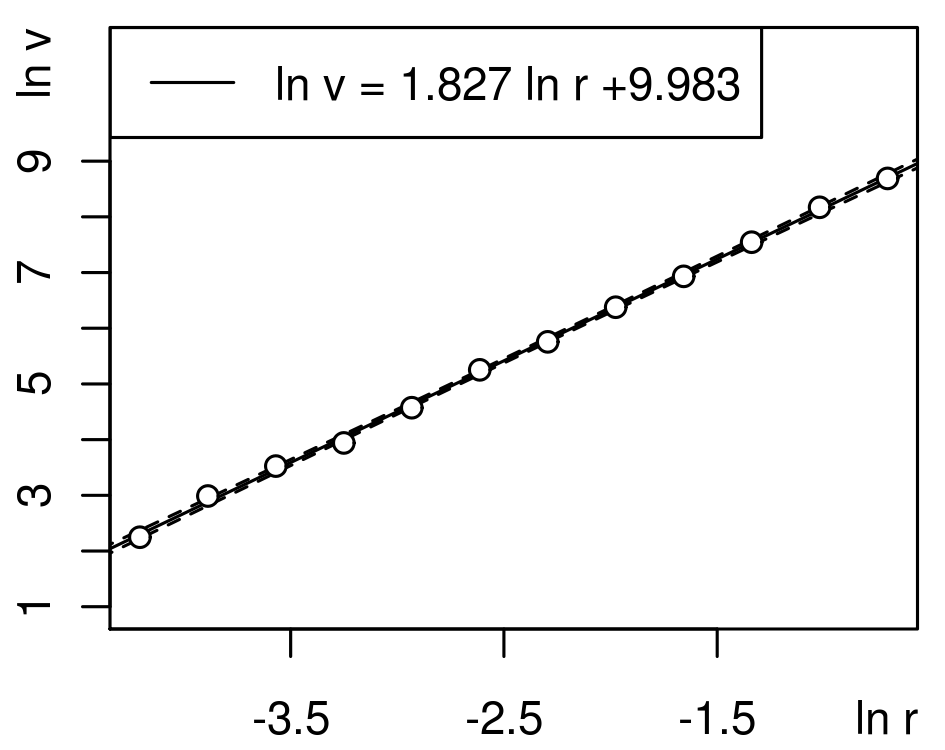}\\[1ex]
\includegraphics[width=.45\linewidth]{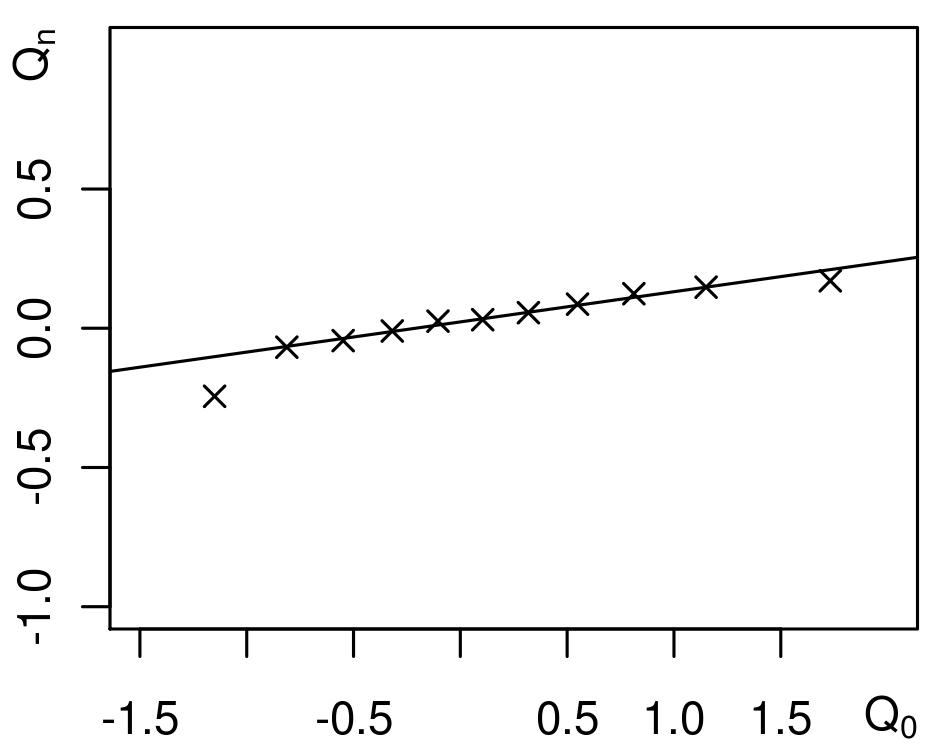}\quad
\includegraphics[width=.45\linewidth]{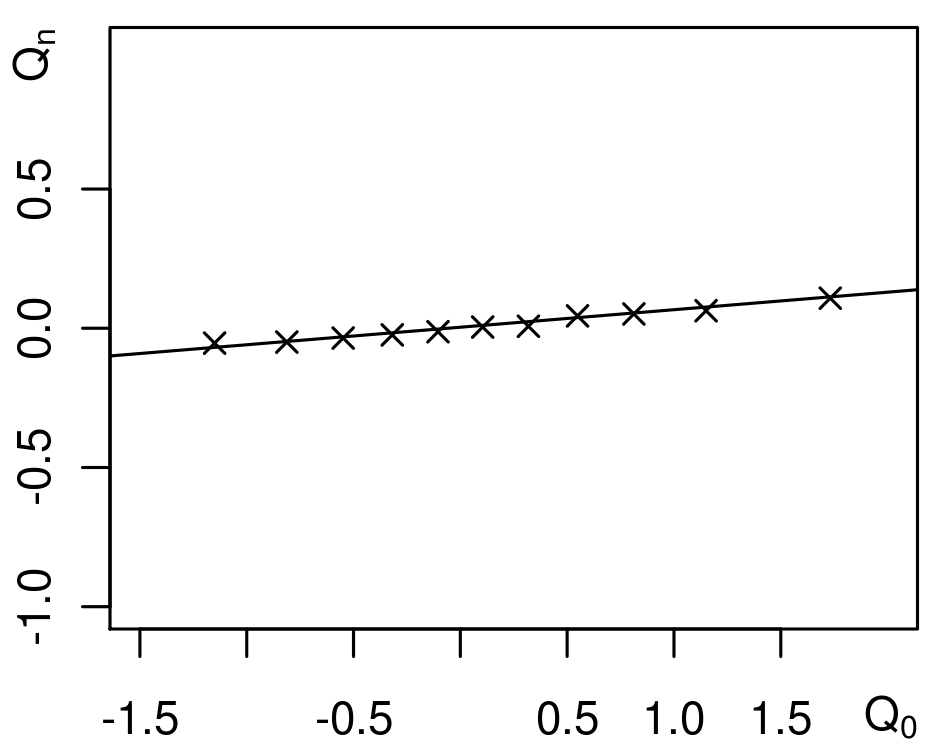}
\caption{Линейные модели (вверху) и Q\--Q графики остатков $e_{bi}$ (внизу) к оценке массовой фрактальной размерности $d_{b1}$ для показанных на рис.\,\ref{pic:samples} выборок кластеров при $p=p_c-0,02$ (слева) и $p=p_c+0,02$ (справа)}
\label{pic:smp.reg}
\end{figure}

Решение задачи линейной регрессии для статистического оценивания массовой фрактальной размерности $d_{b1}$ анализируемых кластеров показано в верхнем ряду на рис.\,\ref{pic:smp.reg}, а соответствующие $p_c-0,02$ и $p_c+0,02$ долям точечные оценки $d_{b1}$ равны $1,397$ и $1,827$ при коэффициентах детерминации $R^2$, чрезвычайно близких к единице: $0,991$ и $0,999$. Символами ``$\circ$'' на рис.\,\ref{pic:smp.reg} показаны эмпирические точки $(\ln r_i, \ln v_i)$, сплошными линиями\--- линии регрессии, а штриховыми\--- их $0,95$\-/доверительные интервалы.

Ещё одним результатом вышеописанного подхода является построение соответствующих $p_c-0,02$ и $p_c+0,02$ долям достижимых узлов $0,95$\-/доверительных интервалов массовой фрактальной размерности $d_{b1}$ выборочных совокупностей $\mathbf{I}_{0,95}(d_{b1})$, которые будут равны $(1,311; 1,484)$ и $(1,791; 1,863)$.

Заметим, что при докритическом значении доли достижимых узлов $p=p_c-0,02$ радиус доверительного интервала по\-/прежнему вдвое больший, а распределение отклонений эмпирических точек от линии регрессии хуже соответствует гипотезе о нормальном распределении ошибок, по сравнению со сверхкритическим режимом $p=p_c+0,02$. Однако, сравнительный анализ Q\--Q графиков остатков $e_{bi}$, приведённых на рис.\,\ref{pic:smp.reg} внизу, показывает, что даже при $p<p_c$ эмпирическое распределение остатков $e_{bi}$ оказывается близким к нормальному, что в целом позволяет говорить о достоверности оценок фрактальной размерности выборок кластеров $d_{b1}$ на всех режимах. Кроме того, из общих соображений ясно, что с ростом доли достижимых узлов $p\to 1$ массовая фрактальная размерность $d_{b1}$ будет возрастать, стремясь к своему предельному значению $d_{b1}\to 2$.

\section{Заключение}

Построение репрезентативных векторов $(\ln r_i, \ln n_i)$ для отдельных реализаций кластеров возможно лишь для критических и сверхкритических долей достижимых узлов $p \geqslant p_c$. Предлагаемый метод использует распределение относительных частот узлов решётки $v_i$ и обеспечивает корректный учёт кластеров малых размеров в выборочной совокупности, что позволяет получать статистические оценки, более адекватные исследуемым явлениям.

Ещё одной важной особенностью предложенного метода является существенное сокращение трудоёмкости вычислений за счёт исключения необходимости решения задачи регрессии для каждой реализации кластера. Вполне очевидно, что указанная методика применима для получения достоверных и вычислительно эффективных оценок большинства статистически устойчивых характеристик перколяционных кластеров.
Все это позволяет говорить о полезности описанного подхода для повышения эффективности применения перколяционных моделей как в теоретических, так и в прикладных исследованиях.

\clearpage

\section*{Сведения об авторах}

\begin{description}
\item[Москалев Павел Валентинович:] кандидат технических наук, доцент кафедры высшей математики и теоретической механики Воронежского государственного аграрного университета имени К.Д. Глинки. \\
Тел.: +7 (473) 253-73-71; e-mail: moskalefff@gmail.com

\item[Гребенников Константин Владимирович:] аспирант кафедры промышленной энергетики Воронежской государственной технологической академии. \\
Тел.: +7 (473) 255-44-66; e-mail: greb86@mail.ru

\item[Шитов Виктор Васильевич:] доктор технических наук, профессор, заведующий кафедрой промышленной энергетики Воронежской государственной технологической академии. \\
Тел.: +7 (473) 255-44-66; e-mail: svw@list.ru
\end{description}

\English
\section*{About the authors}

\begin{description}
\item[Moskalev Pavel Valentinovich:] Ph. D., Associate Professor, Department of Mathematics and Theoretical Mechanics, Voronezh State Agricultural University after K.D. Glinki. \\
Tel.: +7 (473) 253-73-71; e-mail: moskalefff@gmail.com

\item[Grebennikov Konstantin Vladimirovich:] Postgraduate Student, Department of Industrial Power Engineering, Voronezh State Technological Academy. \\
Tel.: +7 (473) 255-44-66; e-mail: greb86@mail.ru

\item[Shitov Viktor Vasiljevich:] Doctor of Engineering Science, Full Professor, Head of Department of Industrial Power Engineering, Voronezh State Technological Academy. \\
Tel.: +7 (473) 255-44-66; e-mail: svw@list.ru
\end{description}

\end{document}